# User office, proposal handling and analysis software

Jörn Beckmann, Jürgen Neuhaus

Technische Universität München
ZWE FRM-II
D-85747 Garching, Germany
jbeckman@frm2.tum.de

At FRM-II the installation of a user office software is under consideration, supporting tasks like proposal handling, beam time allocation, data handling and report creation. Although there are several software systems in use at major facilities, most of them are not portable to other institutes. In this paper the requirements for a modular and extendable user office software are discussed with focus on security related aspects like how to prevent a denial of service attack on fully automated systems. A suitable way seems to be the creation of a web based application using Apache as webserver, MySQL as database system and PHP as scripting language.

## 1      Requirements of user office software

At FRM-II a user office will be set up to deal with all aspects of business with guest scientists. The main part is the handling and review of proposals, allocation of beam time, data storage and retrieval, collection of experimental reports and creation of statistical reports about facility usage. These requirements make off-site access to the user office software necessary, raising some security concerns which will be addressed in chapter 2.
The proposal and beam time allocation process is planned in a way that scientists draw up a short description of the experiment including a review of the scientific background and the impact results from the planned experiment might have. Different users might be grouped together in order to see or modify proposals of their group. The approval of accepting new group members can be delegated to an external head of the group. The proposal is expected to be up to 3 pages long and might include figures and tables. A special layout is not necessary, but the system should be able to handle captions and formulas properly. After internal review of safety and radio protection issues, this proposal is passed on to some referees, who will attach their comments to the proposal and send it back to the user office. The user office should help the user office staff by means of automatic reminders and pass on decisions on the proposal to the submitting scientist, including comments of the referee board. After the proposal is approved, the visiting scientist and the instrument responsible are informed about the allocated beam time. All data taken during the visit is tagged identifying the scientist and the proposal. All data files are stored in a central repository, where only visiting scientists and the instrument responsible have access to. Information about the experiments performed, used sample environment, and the field of research are stored in a central database for statistical purpose. After finishing the experiment the scientist has to prepare a report to be published in the annual reports of FRM-II. The task of the user office software is to collect the reports and to remind scientist about approaching deadlines. After all reports are collected they are processed to be accessible online and to be printed. In contrast to proposals experimental reports require formatted text, footnotes, figures, tables and formulas in higher quality.

The last job of user office software is to deliver data required by the departments responsible for site entry, travel arrangements or reimbursement of travel costs. As this aspect of user office software is site-specific, it is not further investigated in this paper.

## 2    Security issues

The database connected with the user office software holds personal and confidential data which must not be disclosed to unauthorized persons. This require access control for on-site personal and reliable authentication of scientists submitting proposals or requesting access to experimental data. While access control for on-site personal could be done by means of standard operating system features like username/password authentication and access control list in the file system. For off-site access it is not useful to issue a shell login for each visiting scientist and it would require to keep user office logins strictly separated from logins to experimental computers as they might be compromised otherwise. On the other hand issuing two shell logins, one for the user office and one for the experiment doubles the amount of management work.

A solution for this problem is to use a database with username password pairs to be used for access to the user office software. If a user requires a real shell login, for example during the time he is on site doing measurements, the user office software could create and remove the shell account automatically. Another problem is to find a method for checking user particulars before the first login. A relative safe method is to accept first time requests only by mail and send initial password/username by snail-mail. This is on the other hand a slow method and burdens the people form user office with a lot of additional work. A faster way is to accept requests by email, which could be automatically processed. But a faked email address is nothing unusual today. An additional risk of full automatic subscription is that the user office software might be subject of a denial of service attack (DoS), which means that the system gets flooded by lots of registration requests from faked email addresses. When the user office software tries to send the initial username/password information to that false address, it will get a reply about an undeliverable email, which also has to be processed. Finally the whole system is just busy with processing emails and does not longer react to other requests. The easiest way to prevent such situation is to set a limit of requests per minute and to drop all requests in excess to that limit. This of course results in an unpleasant behavior of the whole software. A way to separate handmade requests from automatic produced ones is to create a request form, which will display an information like a word or a number in a picture. To send a valid request, this information has to be included in the request and will change every time the form is accessed. The drawback is, that the system has to spend some effort on keeping track of the required information and it still has to process incoming mails. Only the part of sending information to non-existing email addresses is removed by such complicated procedure.

The last security related problem is to guarantee integrity of transferred information. While information concerning visiting scientists is normally not top secret it is nevertheless annoying to get wrong details about beam time for example. As there are of the shelf methods like SSL/TLS to deal with that the point is not elaborated here. Other requirements like data safety, fault-management and so on, should be a natural course of action and are not part of this paper.

## 3      Implementation details

After the requirements and especially the security related features of a user office software have been discussed a few words about implementing them will follow. We think that a database driven web-based application is ideal to fulfill the requirements. As web-server we chose Apache, whose modular structure provides the flexibility required. Apache is one of the major web servers in use and is well known for its reliability. Together with the TLS module Apache allows encrypted information delivery by means of the HTTPS protocol. The software itself should be implemented in a server-based language like Python, Perl, Tcl or Java. Because of former in-house experience we have chosen PHP as the language to be used. It can be embedded within HTML sourcecode and provides features like database access or on the fly PDF creation. The database should support transactions. Speed is not an issue as the user office software is not expected to handle several requests per second. Commercial database systems like Oracle or DB2 are therefore not required. We decided to use MySQL, again because of former in house experience.

The submission of proposals will be handled by a web form where text could be entered or uploaded. Acceptable formats might be either TeX or plain ASCII, for pictures GIF or JPEG. Accepting proposals as Word or PDF-Files would make the automatic processing more complicated. For the presentation of proposals to the referees, the PDF creation ability of PHP will be used. Experimental reports will also be accepted in PDF or PS form. If requested by users an automated word to PDF/PS conversion feature could be added.

Experiment data will be stored in the NeXus format and a database with meta-information retrieved from the NeXus files will allow to search for datasets. Data integrity and fault tolerance are achieved by using two servers connected to a two-channel RAID-5 subsystem. For disaster tolerance an off site backup using the TSM software from IBM is done once a day.

# 4     Conclusion

User office software is a complex system. Several implementations exist at different facilities. Unfortunately each facility has different requirements and often the user office software is connected strongly to and dependent on the whole in house IT environment. Such a software could hardly be ported to suit the needs of other facilities. A modular software which separates requirements from implementation is the only way to handle this. At FRM-II we therefore decided to interconnect standard software like Apache, PHP and MySQL. It is not our intention to reinvent the wheel again and we therefore happy to accept contributions either in ideas, discussion or software to create a modular and extendable user office software for the new Munich research reactor FRM-II Especially the possible use of software packages like Zope or Midgard is still an open point and we would like to know about experiences made by other facilities with similar requirements.